\documentclass[onecolumn,superscriptaddress,nobalancelastpage,12pt,pra,a4paper]{revtex4}
\usepackage{graphicx}
\usepackage{dcolumn}
\usepackage{bm}
\usepackage{verbatim}
\usepackage{soul}
\usepackage[dvipsnames]{xcolor}
\colorlet{RED}{red}
\usepackage{braket}
\usepackage{dsfont}
\usepackage{amsmath,subfigure,float}
\usepackage{url}
\usepackage{ctable}
\usepackage{soul}
\usepackage[USenglish]{babel}%naustrian
\usepackage{titlesec}
\addto{\captionsUSenglish}{}

\newcommand{\e}{\mathrm{e}}

\renewcommand{\i}{\mathrm{i}}

\titleformat{\section}{\normalfont\large\bfseries}{}{0pt}{}
\titleformat{\subsection}{\normalfont}{}{0pt}{\ul}
\titlespacing{\subsection}{0pt}{\parskip}{4pt}
\titleformat{\subsubsection}{\normalfont\large}{}{0pt}{}
\titlespacing{\subsubsection}{0pt}{\parskip}{4pt}

\usepackage{xr}

\makeatletter
\newcommand*{\addFileDependency}[1]{% argument=file name and extension
  \typeout{(#1)}
  \@addtofilelist{#1}
  \IfFileExists{#1}{}{\typeout{No file #1.}}
}
\makeatother

\newcommand*{\myexternaldocument}[1]{%
    \externaldocument{#1}%
    \addFileDependency{#1.tex}%
    \addFileDependency{#1.aux}%
}

\myexternaldocument{SM}

\begin{document}

\title{Quantum holography with undetected light}
\author{Sebastian T{\"o}pfer$^{1,\dagger}$ , Marta Gilaberte Basset$^{1,2,\dagger}$, Jorge Fuenzalida$^{1}$, Fabian Steinlechner$^{1,2}$, Juan P. Torres$^{3}$, and Markus Gr{\"a}fe$^{1,2,*}$
\vspace{1em}
\\
\it $^{1}$Fraunhofer Institute for Applied Optics and Precision Engineering IOF,
\it Albert-Einstein-Str. 7, 07745, Jena, Germany. \\ \vspace{0.5em}
\it $^{2}$Friedrich-Schiller-University Jena, Institue of Applied Physics, Abbe Center of Photonics,\\ 
\it Albert-Einstein-Str. 6, 07745, Jena, Germany.\\ \vspace{0.5em}
\it $^{3}$ICFO-Institut de Ciencies Fotoniques, The Barcelona Institute of Science and Technology,\\
\it 08860 Castelldefels, Spain, and Dept. Signal Theory and Communications, Universitat Politecnica de Catalunya, 08034 Barcelona, Spain.\\ \vspace{0.5em}
$^{*}$corresponding author. Email: markus.graefe@iof.fraunhofer.de\\
$^{\dagger}$Both authors contributed equally.
}
\date{\today}

%\begin{abstract}
%Holography exploits the interference of light fields to obtain a systematic reconstruction of the light fields wavefronts. Classical holography techniques have been very successful in diverse areas such as microscopy, manufacturing technology, and basic science.
%, to name a few. %
%Extending holographic methods to the level of single photons has been  proven challenging, since applying classical holography techniques to this regime pose technical problems. Recently the retrieval of the spatial structure of a single photon, using another photon under experimental control with a well-characterized spatial shape as reference, was demonstrated using the intrinsically non-classical Hong-Ou-Mandel interference on a beam splitter%~\cite{Chrapkiewicz:2016}
%. Here we present a method for recording a hologram of  single photons without detecting the photons themselves, and importantly, with no need to use a well-characterized companion reference photon. Our approach is based on quantum interference between two-photon probability amplitudes in a nonlinear interferometer%~\cite{Mandel:1991,Lemos:2014}
%. As in classical holography, the hologram of a single photon allows retrieving the complete information about the "shape" of the photon (amplitude and phase) despite the fact that the photon is never detected. 
%\end{abstract}

\maketitle

\newpage
%---------------------ABSTRACT-----------------------
\section{Abstract}
%Holography exploits the interference of light fields to obtain a systematic reconstruction of the light fields wavefronts. Classical holography techniques have been very successful in diverse areas such as microscopy, manufacturing technology, and basic science. Extending holographic methods to the level of single photons has been proven challenging, since applying classical holography techniques to this regime pose technical problems. Recently the retrieval of the spatial structure of a single photon, using another photon under experimental control with a well-characterized spatial shape as reference, was demonstrated using the intrinsically non-classical Hong-Ou-Mandel interference on a beam splitter. Here we present a method for recording a hologram of  single photons without detecting the photons themselves, and importantly, with no need to use a well-characterized companion reference photon. Our approach is based on quantum interference between two-photon probability amplitudes in a nonlinear interferometer. As in classical holography, the hologram of a single photon allows retrieving the complete information about the "shape" of the photon (amplitude and phase) despite the fact that the photon is never detected. 

%Shorted version for Science Advances - max 150 words
Holography exploits the interference of a light field reflected/transmitted from an object with a reference beam to obtain a reconstruction of the spatial shape of the object. Classical holography techniques have been very successful in diverse areas such as microscopy, manufacturing technology, and basic science. However, detection constraints for wavelengths outside the visible range restrict the applications of holography techniques for imaging and sensing in general.  %Extending holographic methods to the level of single photons has been proven challenging, since applying classical holography techniques to this regime pose technical problems. Recently the retrieval of the spatial structure of a single photon, was demonstrated using the intrinsically non-classical Hong-Ou-Mandel interference on a beam splitter.
For overcoming such detection limitations, we implement phase shifting holography with nonclassical states of light, where we exploit quantum interference between two-photon probability amplitudes in a nonlinear interferometer. We demonstrate that it allows retrieving the spatial shape (amplitude and phase) of the photons transmitted/reflected from the object, and thus obtaining an image of the object, despite those photons are never detected. Moreover, there is no need to use a well-characterized reference beam, since the two-photon scheme already makes use of one of the photons as reference for holography.   %It allows retrieving the complete information about the "shape" of the photon (amplitude and phase) despite the fact that the photon is never detected. 

% Teaset for Science Advances - max 125 characters
\section{Teaser}
Experiments show the performance of quantum holography using undetected photons.

\newpage

%---------------------INTRO-----------------------
\section{Introduction}
In the last few decades, scientists and engineers throughout the world, from different disciplines, governments and information technology companies are paying increasing attention to quantum technologies. Quantum communications~\cite{Xu:2020,Wengerowsky:20,ortega:2021}, quantum computation~\cite{zhong:2020}, and in particular quantum imaging~\cite{brida:2010,Lemos:2014,Kalashnikov:2016,Gilaberte:2019,moreau:2019}, %gregory:2020} 
are just some examples of novel areas of science and technology where quantum ideas are helping to implement systems with enabling new capabilities. Quantum technologies promise to go further than classical counterpart technologies by using new quantum states of light and matter, performing tasks that are impossible to implement classically~\cite{defienne:2021}. 
%A clear example of this is the ability of obtaining information of samples without the direct detection of the light that interacts with it \cite{Lemos:2014,Kalashnikov:2016, Paterova:2018, Valles:2018, Kutas:2020, Vanselow:2020,Kviatkovsky:2020,Gilaberte:2021}. So far, different degrees of freedom of the light interacting with the sample can be obtained, but, to date, no holograms have been recorded.
A clear example of this is the ability of obtaining a hologram from single photons~\cite{Chrapkiewicz:2016}, and even recording the hologram without detecting the photons themselves as we report here. 

Holography was introduced by D.~Gabor in 1948~\cite{gabor:1948}. It allows the reconstruction of the spatial structure of an object by recording amplitude and phase information of the light reflected from an object. Classical holography is already successfully applied in sensing and microscopy. This is true especially for bio-specimen, where scattering and absorption require a phase-sensitive sensing contrast agent-free approach~\cite{Popescu:2011, Marquet:05}. Holography can also be applied in optical security ~\cite{Refregier:95, LIU2014327} and data storage ~\cite{https://doi.org/10.1002/anie.201002085,10.1117/12.506055}. The introduction of single photon holographic methods would expand holography to several applications that raised together with the recent growth of quantum technologies~\cite{Chrapkiewicz:2016}. In particular, single photon holography with undetected light in a nonlinear interferometer scheme \cite{Yurke:1986,Mandel:1991_2,herzog1994frustrated,Chekhova:16} would introduce the benefit of choosing the most convenient spectral range for the probing beam in an application without facing the limitations imposed by the low efficiency of detectors at those specific spectral range (e.g. in the mid-infrared). Nonlinear interferometry has been proven to be key elements in numerous applications, namely, in imaging~\cite{Lemos:2014,Cardoso:2018,Gilaberte:2021}, sensing~\cite{Kutas:2020}, spectroscopy~\cite{Kalashnikov:2016,Paterova:2018}, microscopy~\cite{Kviatkovsky:2020,Paterova:2020}, and optical coherence tomography~\cite{Valles:2018,paterovaOCT2018,ramelowOCT2020,gerardOCT2020}.
\par
%As a versatile method, holography is successfully applied in sensing and microscopy. Especially for bio-specimen, where scattering and absorption require a phase sensitive sensing contrast agent-free approach~\cite{Popescu:2011, Marquet:05}. Holography can also be applied in optical security ~\cite{Refregier:95, LIU2014327} and data storage ~\cite{https://doi.org/10.1002/anie.201002085,10.1117/12.506055}. Furthermore, 
%
%The essential property of light that enables holography is interference, and thus, its coherence, which is rooted in the heart of quantum physics~\cite{Glauber:1963}. Hence, it is of particular interest to observe holographic phenomena on a quantum level. Recording a hologram of a single photon was believed to be impossible until just recently Chrapkiewicz and co-workers proofed differently~\cite{Chrapkiewicz:2016}. In their seminal work, they utilized two-photon interference on a beamsplitter, referencing one photon with a second one.
%With our technique, we let the photon interfere with itself in contrast to the previous work from Chrapkiewicz and co-workers, and therefore, we expand holography methods to single photons process by combining the concept of induced coherence~\cite{Mandel:1991, Mandel:1991_2, herzog1994frustrated} with digital phase shift holography. This allows us to perform a hologram of a single photon without even detecting the photon itself.
%It allows us to experimentally verify a quadratic phase originating from spontaneous parametric down conversion processes. 
To characterize the spatial shape of the single-photon, we use phase shifting holography, introduced in 1997 by Yamaguichi \& Zhang~\cite{Yamaguchi:1997}, where several images with different phase steps are recorded and processed. This technique allows the reconstruction of the spatial structure (amplitude and phase) of the light reflected from a sample. In combination with the use of the effect of induced coherence~\cite{Mandel:1991_2, herzog1994frustrated} 
%, which enables the so-called quantum imaging with undetected photons~\cite{Lemos:2014, Valles:2018, Paterova:2020, Kviatkovsky:2020,Fuenzalida:2020,Kutas:2020,Gilaberte:2021} 
we can obtain an hologram without detecting the light that illuminated the sample. Our experiment is a step forward to enable efficient holography in a broader spectral range.% and thus paves the way to new photonic based analysis tools.

\section{Results}
\subsection{Phase-shifting holography with a SU(1,1) nonlinear interferometer}
In this part we give a brief overview of the theoretical framework of phase shifting holography in the quantum regime considered here. An extended and more detailed theoretical description of the experiment can be found in the Supplemental Material.

\begin{figure}
	\centering
		\includegraphics[width=0.8\linewidth]{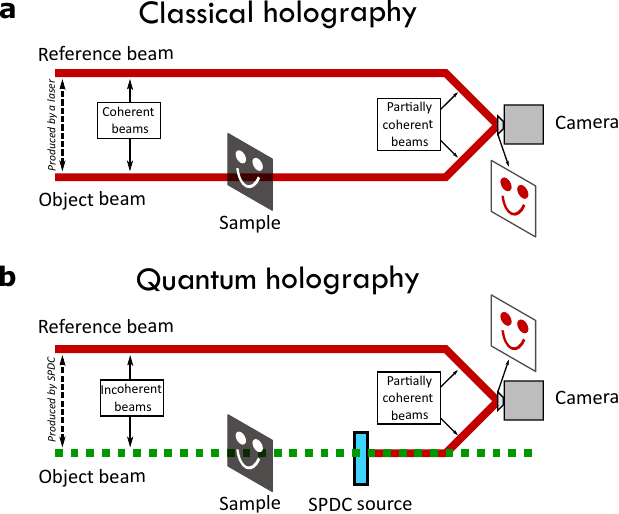}
	\caption{\textbf{Classical and quantum holography.
	} \textbf{a}) In classical holography the spatially-dependent interference pattern of two coherent beams, the reference beam and the object beam, after interaction with the object, are recorded and used to construct the hologram of the object. \textbf{b}) In our quantum holography scheme we make use of two-photon states that can be generated by spontaneous parametric down conversion (SPDC) in one of two sources. The spatial shape of the object, which is transferred to the spatial shape of the light reflected/transmitted from the object, is contained in the probability amplitudes corresponding to the paired photons being generated in either of two SPDC sources.}
	\label{fig:1_schemeidea}
\end{figure}

In classical holography (see Fig.~\ref{fig:1_schemeidea}(a)) the hologram is the result of the interference of two input beams with mutual coherence~\cite{goodman_fourier}. One beam serves as reference beam. The other beam (object beam) illuminates an object and generates reflected/transmitted light that bears information of the spatial structure of the object.  The spatial shape of the interference pattern, resulting from combining the object and reference beams 
\begin{equation}
    I(x,y)=I_{\mathrm{r}}(x,y) +I_{\mathrm{o}}(x,y)+2\sqrt{I_{\mathrm{r}}(x,y) I_{\mathrm{o}}(x,y)} \cos \left[\theta_{\mathrm{o}}(x,y)- \theta_{\mathrm{r}}(x,y) \right]
    \label{classical_holography}
\end{equation}
is recorded in the hologram, and can be visualized with the help of a CCD camera~\cite{goodman_fourier}. Here $I_{\mathrm{r}}$ and $\theta_{\mathrm{r}}(x,y)$ are the intensity and phase of the reference beam, and $I_{\mathrm{o}}$ and $\theta_{\mathrm{o}}(x,y)$ are the intensity and phase of the object beam after reflection/transmission by the object.

For the sake of clarity, let us consider quantum holography with undetected light in an induced coherence scenario (see Fig.~\ref{fig:1_schemeidea}(b)), where reference and object beams are generated by means of spontaneous parametric down-conversion (SPDC). The first important difference with classical holography is that the object beam, after reflection/transmission from the object, is not made to interfere with the reference beam.  The object beam remains \emph{undetected}. 
%The second important difference with classical holography is that the two beams that indeed interfere to register the hologram are not mutually coherent, but partially coherent beams. 
The second important difference for quantum holography is that the reference and object beam are incoherent beams. This can be seen from the fact that object and reference beam in the quantum holography scheme in principle have different spectra, which is indicated by the different colors used in Fig.~\ref{fig:1_schemeidea}(b).
If the transmission function of the object to be recorded is $\tau(x,y)=t(x,y)\e^{\i\theta(x,y)}$ ($|\tau|=t$), the spatially-dependent degree of first-order coherence $g^{(1)}(x,y)$ between the two beams is $g^{(1)}(x,y)=t(x,y)$ (for further details see the information in the Supplementary Material).

In our work, we implement the two beams for the quantum holography scheme with undetected light by a photon pair source based on spontaneous parametric down conversion (SPDC) in nonlinear crystals. There, correlated signal and idler beams are generated. Arranging such two SPDC sources in a nonlinear SU(1,1) interferometer, signal and idler pairs are generated in either of the nonlinear crystals.
Our experiments work in the low parametric gain regime, where the probability to generate signal-idler pairs in both nonlinear crystals simultaneously is negligible. In this regime, one can think that the hologram is the result of quantum interference between two possibilities characterized by
the corresponding probability amplitudes. One possibility is that signal-idler pairs are generated in SPDC, the idler photons interact with the object and impinges on the second nonlinear crystal without inducing the emission of new paired photons. The second possibility is that signal-idler pairs are generated in the second nonlinear crystal, so no idler photons have traversed the object. The transmission function of the object determines the distinguishability of both possibilities and thus the degree of coherence of the two beams that interference in the hologram.

For quantum holography with undetected light one finds (see Supplementary material for a detailed derivation) that the spatially-dependent flux-rate of signal photons detected $N_{\Delta\varphi}(x,y)$ when an object with transmission function $t(x,y)\e^{\i\theta(x,y)}$ is present in the idler path is~\cite{Gilaberte:2021}
\begin{equation}
    N_{\Delta\varphi}(x,y)\sim1+t(x,y)\cos \left[ \theta(x,y)-\nu(x,y)+\Delta\varphi \right],
\end{equation}
where $\Delta \varphi$ is a global phases and $\nu(x,y)$ is a phase introduced by the nonlinear parametric down-conversion processes inside the nonlinear interferometer. This equation is slightly different from Eq.~\eqref{classical_holography} that describes the interference pattern registered in classical holography. Nevertheless, both equations show that the amplitude and phase of light reflected/transmitted by an object can be registered in a medium (hologram) sensitive only to the intensity of light. 

We aim at visualizing the interference pattern to be recorded in the hologram. For an unknown object a single measurement does not allow to extract full phase ($\theta(x,y)$) and amplitude ($t(x,y)$) information of the complex transmission coefficient introduced by the object. However, phase shifting holography can be applied in order to extract such information, both amplitude and phase. In doing so, a series of images with global phases $\Delta\varphi=0,\pi/2,\pi,3\pi/2$ can be recorded and processed.  This will result in four images:
\begin{eqnarray}
  N_{0}(x,y)  &\sim&  1+t(x,y)\cos\theta(x,y) \nonumber\\
  N_{\pi/2}(x,y) &\sim& 1-t(x,y)\sin\theta(x,y) \nonumber\\
  N_{\pi}(x,y) &\sim& 1-t(x,y)\cos\theta(x,y)\nonumber\\
  N_{3\pi/2}(x,y) &\sim& 1+t(x,y) \sin\theta(x,y)
    \label{equ:DPSH_example_with_4}
\end{eqnarray}
From these four measurements one can easily extract the phase information
\begin{equation}
   \theta(x,y) = \arctan\left( \frac{N_{3\pi/2} - N_{\pi/2}}{N_0 - N_{\pi}} \right),
    \label{equ:DPSH_example_with_4_phase}
\end{equation}
and the amplitude information
\begin{equation}
   t(x,y) = 2 \times \frac{\left\{ \left[ N_{3\pi/2} - N_{\pi/2} \right]^2 + \left[ N_0 - N_{\pi} \right]^2 \right\}^{1/2}}{N_0 + N_{\pi/2} + N_{\pi} + N_{3\pi/2}}.
    \label{equ:DPSH_example_with_4_amplitude}
\end{equation}
Recording four images of the spatially-varying signal photon flux rate at a certain spectral range with a high efficiency detector allows to retrieve full phase and amplitude information of an unknown object that is probed by photons in a different spectral range. This way quantum holography with undetected light becomes possible.
\par
One can generalize this approach to a series of $M\geq3$ images~\cite{malacara:2007}. They need to be recorded with phases $\Delta\varphi_m=2m\pi/M$ with $m=0\ldots M-1$. Then, phase and amplitude information are given by
\begin{equation}
    \theta(x,y) = -\arctan \left( \frac{\sum_m N_{\Delta\varphi_m}(x,y)\sin\Delta\varphi_m}{\sum_m N_{\Delta\varphi_m}(x,y) \cos\Delta\varphi_m}\right)
    \label{equ:general_phase}\\
\end{equation}   
and
\begin{equation}
    t(x,y) = 2 \times \frac{\left\{\left[\sum_m N_{\Delta\varphi_m}(x,y)\cos\Delta\varphi_m\right]^2 + \left[\sum_m N_{\Delta\varphi_m}(x,y)\sin\Delta\varphi_m\right]^2\right\}^{1/2}}{\sum_m N_{\Delta\varphi_m}(x,y)}.
    \label{equ:general_amplitude}
\end{equation}
One would expect that the consideration of a higher number of steps $M$ will lead to a more accurate and precise reconstruction of the phase and amplitude as it is the case in classical phase shifting holography. Consequently, we have experimentally implemented quantum phase shifting holography with undetected light for different step numbers. We have analyzed their impact on accuracy as well as the overall performance of each approach. \\
%------------------------------------------------------

\subsection{Experimental setup}
Most nonlinear interferometers implemented so far for quantum imaging and spectroscopy make use of one of two configurations. In one configuration the signal photons generated in the first nonlinear crystal are detected and they never traverse the second nonlinear crystal~\cite{Mandel:1991_2,Lemos:2014,Valles:2018}. Only idler photons generated in the first nonlinear crystal, after being reflected/transmitted by the object impinge on the second nonlinear crystal. This is the original configuration put forward in 1991 by L. Mandel's group~\cite{Mandel:1991,Mandel:1991_2}.

In an alternative configuration, usually termed as SU(1,1) interferometer, the signal photons generated in the first nonlinear crystal are also injected in the second nonlinear crystal as the idler photons~\cite{Yurke:1986,herzog1994frustrated,Paterova:2018,Gilaberte:2021,gerardOCT2020}. Quantum interference explains the physical origin of the interference pattern for both configurations, and the mathematical expressions that describe the shape of this interference pattern are mostly equal. However, in the original configuration considered we have three beams at the output (two signal beams and one idler beam), while in the SU(1,1) configuration the output consists of two beams (signal and idler beams)~\cite{gerardOCT2020}.

\begin{figure}[tb]
	\centering
		\includegraphics[width=\linewidth]{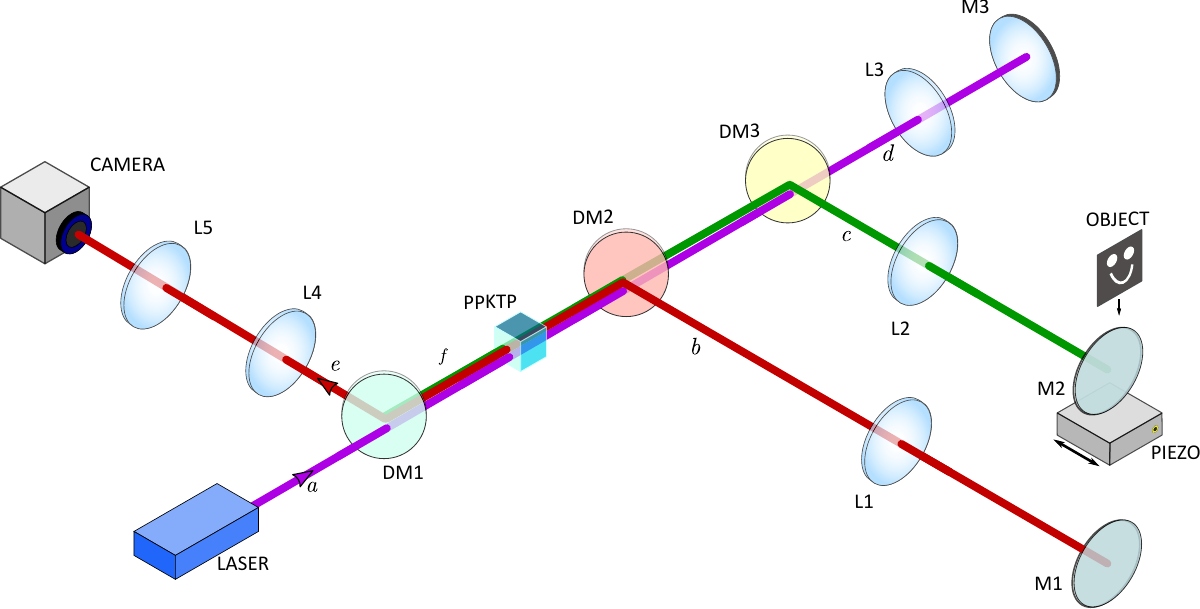}
	\caption{\textbf{Experimental setup for holography with undetected light.} Laser light (purple) pumps the nonlinear crystal (ppKTP) bi-directionally (beam paths $\mathrm{a}$ and $\mathrm{d}$). It generates signal (red) and idler (green) beams either in the forward direction (beam paths $\mathrm{b,c}$) or backward direction (beam paths $\mathrm{e,f}$). Dichroic mirrors DM1 to DM3 separate the different beam paths. Idler light will illuminate the object (beam path $\mathrm{c}$), while its hologram will be detected on the sCMOS camera with the signal light (beam path $\mathrm{e}$). The mirrors M1 to M3 are the interferometer end mirrors. M2 is mounted on a piezo stage to precisely move the mirror in one direction. Lenses L1 to L5 form the imaging system with the focal distances of 150~mm (L1, L2, L3), 100~mm (L4), and 125~mm (L5).
	}
	\label{fig:2_setup}
\end{figure}

Our experimental implementation of quantum holography with undetected light makes use of an SU(1,1) interferometer in a Michelson geometry as shown in Fig.~\ref{fig:2_setup}. It consists of a nonlinear periodically-poled potassium titanyl phosphate (ppKTP) crystal of 2~mm length. The crystal is pumped bi-directionally with a collimated 405 nm cw laser beam with up to $90$ mW of pump power. Correlated signal and idler beams are emitted from the crystal via SPDC with central wavelengths of 910 nm and 730 nm, respectively. As shown in Fig.~\ref{fig:2_setup}, the crystal is imaged onto itself with a 4f-system in each of the three interferometer arms via lenses L1 to L3. The object is placed in the Fourier plane of the idler arm having the momentum space at the object location. The interferometer end mirror M2 in the idler arm is mounted on a piezo translation stage~\cite{smarAct:2021} that allows to scan different phases of the interference produced by varying the path length. This way, one can define specific positions for the mirror M2 corresponding to different phase values $\Delta\varphi$. There is quantum interference between the two possibilities for the generation of signal-idler paired photons, after the pump, idler, and signal beams return to the crystal. In this way the object's information imprinted on the idler light is transferred  to a spatially-varying intensity of the signal light. Lenses L4 and L5 are used to form an image of the object at the camera plane. The camera~\cite{photometrics:2021} has $2048\times2048$~px with a pixel size of 6.5~$\mu$m. The images shown in this work are $500\times500$~px in size. The total usable field-of-view has a diameter of $\approx$ 6.1~mm.\\
%-----------------------------------------------
\subsection{Holography and Imaging performance}
%To experimentally test the quantum holography with undetected light approach we used objects that were generated by direct writing grayscale lithography. A substrate is coated with a photoresist of an index by 1.63 (520nm) and exposed to UV-light in a particular pattern. The chemical solubility of the exposed areas increases in alkaline media and the microstructure in the photoresist is formed.
To experimentally test the quantum holography with undetected light approach we used objects that were engraved in glass plates of refractive index 1.6 by grayscale lithography. The height of the engraved area was designed to induce a total phase change on the idler beam of either $0.62\pi$ or $0.82\pi$, depending on the object. The dimensions of the objects and their features are shown in Figure \ref{fig:X_object_dimensions} in the Supplemental Material.
\par
\begin{figure}[tb]
	\centering
		\includegraphics[width=0.95\linewidth]{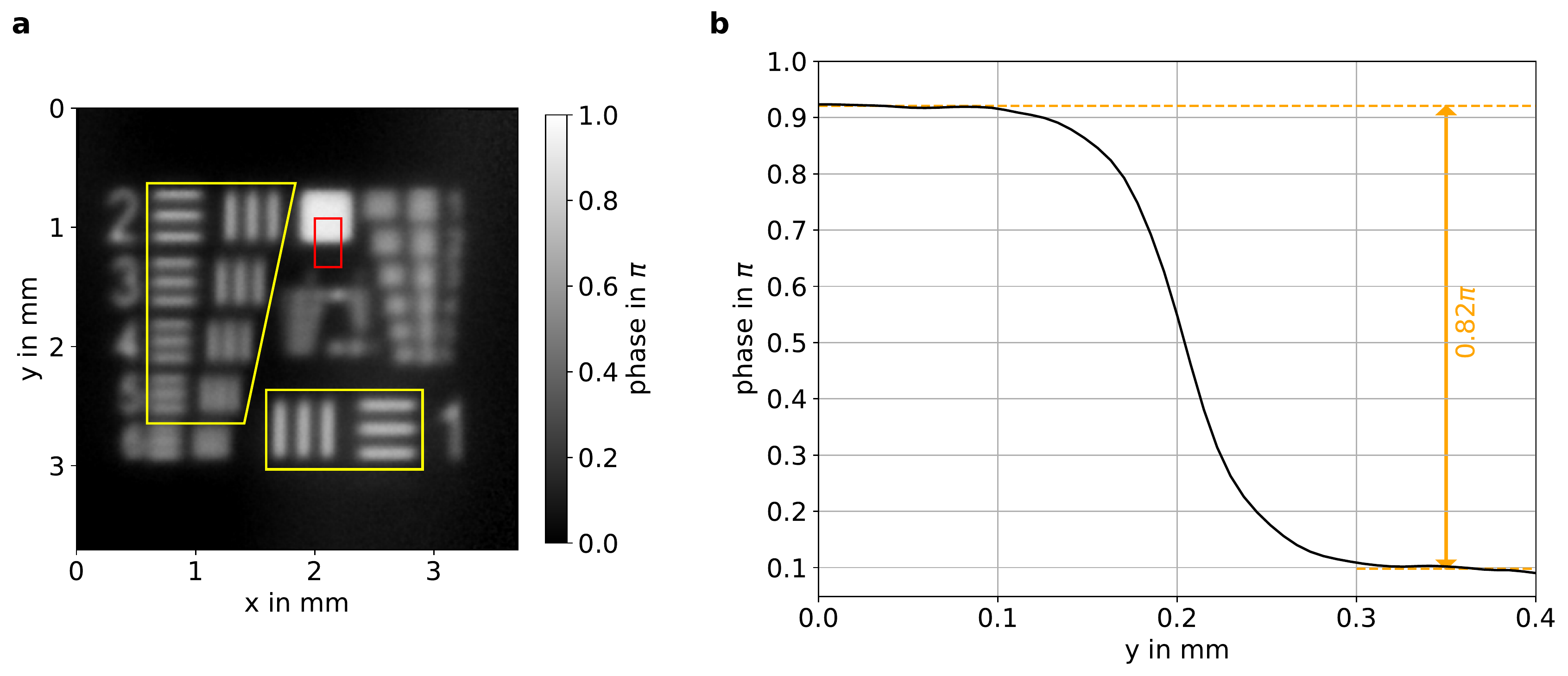}
	\caption{\textbf{Hologram with undetected light of a resolution target.} (a) Wide field holographic image (using 12 frames) of the miniaturized resolution target with phase step of $0.82\pi$ (at the illumination wavelength). The elements contained in the yellow marked areas are the ones analysed to determine the resolution of the setup (see Tab.~ \ref{tab:x_bar_resolution}). The red rectangle highlights the area used to verify the phase step of the object. (b) Phase step plot of the object, which matches with the manufactured $0.82\pi$ value.
	\label{fig:X_phase_image_and_step}
	}
\end{figure}

Two of the objects are resolution targets that are a miniaturized version of the standard 1951 USAF resolution target in order to make all structures fit inside the field of view of the system (6.1 mm in diameter). One is with $0.62\pi$ phase step, the other with $0.82\pi$. A full wide field holographic image for the latter one can be seen in Fig.~\ref{fig:X_phase_image_and_step}(a). The area marked with a red rectangle is used to evaluate the phase step. As exemplarily shown in Fig.~\ref{fig:X_phase_image_and_step}(b) the desired phase step of $0.82\pi$ is well matched. 

\begin{figure}[H]
	\centering
		\includegraphics[width=0.8\linewidth]{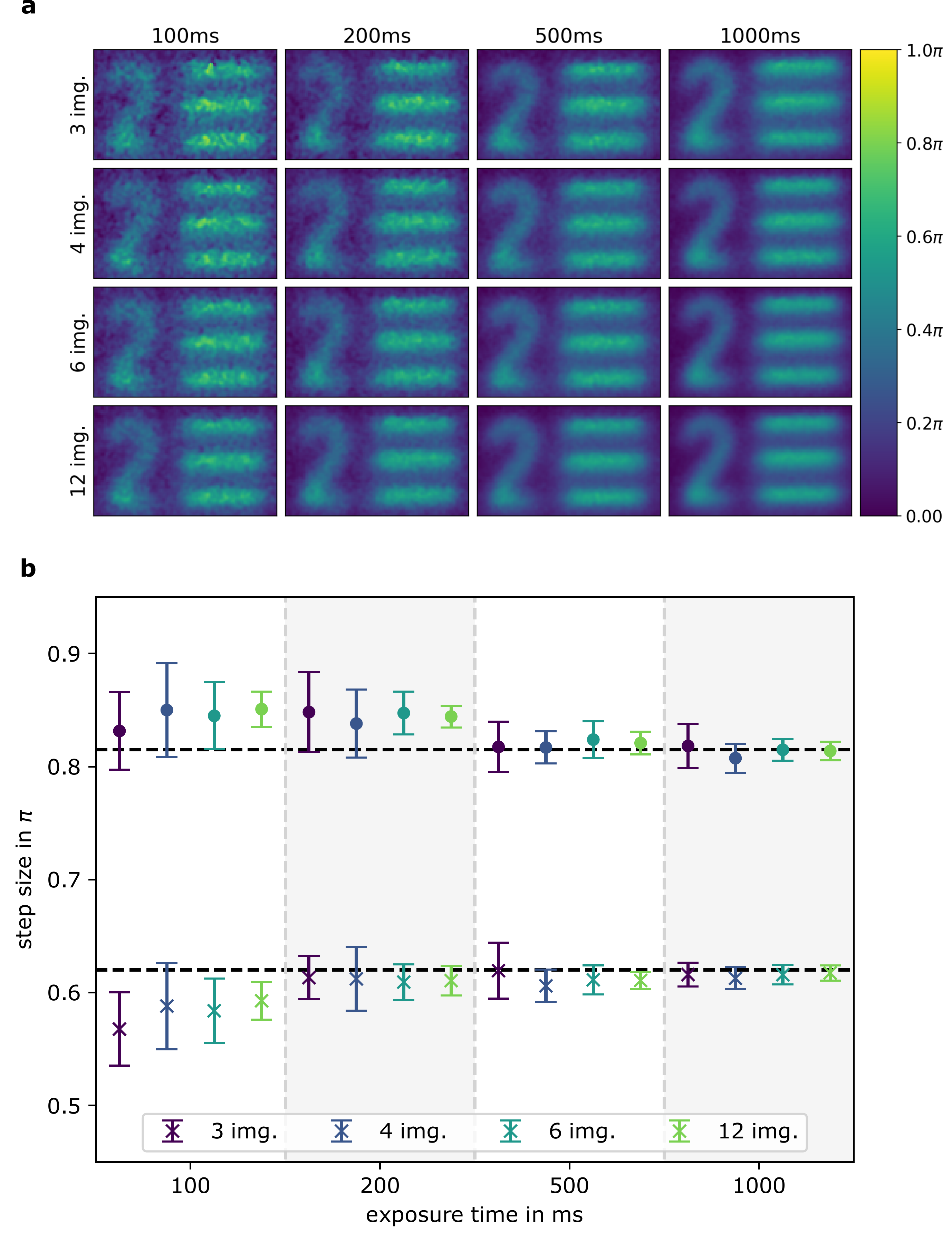}
	\caption{\textbf{Phase accuracy.} (a) Images of the same structure for different acquisition times of the camera and different number $M$ of images used for the phase shifting holography. (b)
	Calculated phase steps. Each point is an average of 15 image sets. The black dashed lines are the expected results. The color is in relation to the amount of images used (see legend). The cross markers refer to a sample with a step size of $0.62\pi$ and the bullet markers refer to a sample with $0.82\pi$ step.
	\label{fig:X_phase_image_matrix}
	}
\end{figure}

With both resolution objects, we investigated the impact of the number of images $M$ recorded for the phase shifting holography, as well as the acquisition time per shot on a resolution target. In doing so, we varied the number of images from three to twelve and the acquisition time per image from 100 ms to 1 s. The results are shown in Fig.~\ref{fig:X_phase_image_matrix}. A four by four image matrix shows the region of interest assigned to the 'bar pattern 2' (Fig.~\ref{fig:X_phase_image_and_step}) under different measurement settings. It clearly indicates that a higher number of images for the phase shifting holography as well as a higher acquisition time lead to a better image quality. A quantitative evaluation is displayed in Fig.~\ref{fig:X_phase_image_matrix}(b) for both resolution targets. 
The results of the phase step retrieval show that the calculated phase values match the expected values for exposure times higher than 200 ms. An increase in the amount of images or exposure time lowers the standard deviation of the results, due to a decrease of the measured noise (see Fig.~\ref{fig:X_phase_noise} in the Supplemental Material). According to this results, the current setup should be operated with 500 ms exposure time to ensure, that the measured values align with the expected values within the standard deviation. It is possible to do so with only $M=3$ images, but $M=4$ images lead to significant more accuracy. This results in at least 2 s overall measurement time for one phase shifting holography measurement, disregarding the time needed to change the phase position and the post processing of the data. For this case, the maximum phase noise was measured to be $(0.091\pm0.005)\pi$, which is the lowest detectable phase difference. The transmission of the object was measured to be homogeneous with $t=(94\pm1)\%$ as in agreement with the theory taking into account the visibility with and without object~\cite{Gilaberte:2021}. %The visibility of the system, considering component and alignment losses, is $75\pm2\%$ ~\cite{Gilaberte:2021}. When placing the phase object inside, it drops to 67\%, which leads to a $94\pm1\%$ transmission value for the object.
%Figure XX shows the DPSH images from two of the objects tested with the 4 step algorithm and 500ms integration time, which showed the best trade-off between performance and speed (see \ref{fig:X_phase_step_eval}).

The blurring in the phase images shows, that the spatial resolution is visibly limited. Using the yellow marked bar patterns on the resolution target (see Fig.~\ref{fig:X_phase_image_and_step}(a)) the contrast for different spatial frequencies can be calculated. It must be considered that it is a miniaturized version of the standard USAF target. Table~\ref{tab:x_bar_resolution} shows the sizes of the bar patterns and the measured contrast. Using the Rayleigh criterion, the measured contrast must be at least 14.2\% to count as resolvable. Because the contrast is measured using bar patterns, which have multiple frequency components instead of sine patterns, it must be corrected by multiplying it with $\pi/4$~\cite{Nill:2001}. These corrected contrast values are shown in the last two columns, named as sine wave contrast. The lowest measured resolvable spatial frequency according to the Rayleigh criteria is 6.3~mm$^{-1}$, which corresponds to 79~$\mu$m feature size.

\begin{table}[H]
    \centering
    \begin{tabular}{|c|c||c|c||c|c|}
        \hline
        line pair & frequency & \multicolumn{2}{c||}{square wave contrast} & \multicolumn{2}{c|}{sine wave contrast}\\
        in mm & in 1/mm & vertical & horizontal & vertical & horizontal \\
        \hline
        0.200 & 5.0 & 0.60 $\pm$ 0.06 & 0.62 $\pm$ 0.06 & 0.47 $\pm$ 0.06 & 0.48 $\pm$ 0.06\\
        0.178 & 5.6 & 0.53 $\pm$ 0.07 & 0.45 $\pm$ 0.08 & 0.42 $\pm$ 0.06 & 0.35 $\pm$ 0.07\\
        0.158 & 6.3 & 0.36 $\pm$ 0.09 & 0.28 $\pm$ 0.09 & 0.29 $\pm$ 0.08 & 0.22 $\pm$ 0.08\\
        \textcolor{gray}{0.140} & \textcolor{gray}{7.1} & \textcolor{gray}{0.18 $\pm$ 0.10} & \textcolor{gray}{0.12 $\pm$ 0.10} & \textcolor{gray}{0.14 $\pm$ 0.09} & \textcolor{gray}{0.09 $\pm$ 0.09}\\
        \textcolor{gray}{0.125} & \textcolor{gray}{8.0} & \textcolor{gray}{0.07 $\pm$ 0.10} & \textcolor{gray}{---} & \textcolor{gray}{0.06 $\pm$ 0.09} & \textcolor{gray}{---}\\
        \hline
    \end{tabular}
    \caption{\textbf{Spatial Resolution.} Calculated Contrast using the bar patterns of the miniaturized resolution target (yellow marked area in Fig. \ref{fig:X_phase_image_and_step}) for 1000 ms acquisition time and $0.82\pi$ phase step. Square wave contrast is the measured contrast on the bar pattern and sine wave contrast is the corrected result, taken into account that the bar pattern contains multiple frequencies. The rows in gray do not fulfill the Rayleigh criterion.}
    \label{tab:x_bar_resolution}
\end{table}

\begin{figure}[tb]
	\centering
		\includegraphics[width=0.7\linewidth]{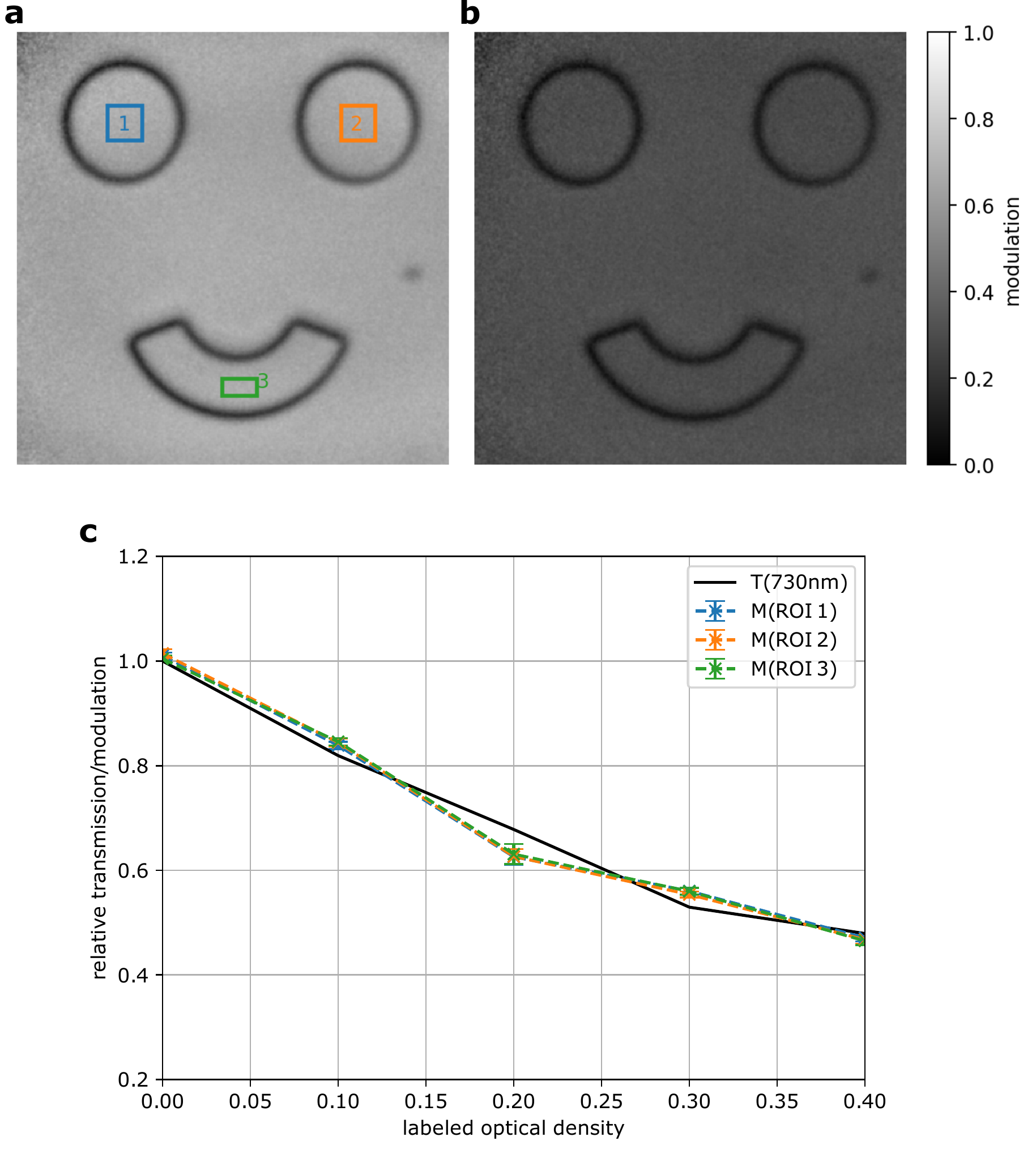}
	\caption{\textbf{Amplitude accuracy.} (a) Modulation image of a $0.82\pi$ phase step mask without optical density filters. (b) Modulation image of the same mask fully covered with an optical density filter (OD 0.4). (c) Relative modulation calculated values for the three different areas marked in the correspondent colors in (a) and expected transmission (black solid line) for different object transmission values, experimentally implemented by placing different optical density filters in front of the phase mask. The modulation values are retrieved from holographic images with $M=12$ phase steps each with 500~ms exposure time.
	\label{fig:X_mod_vs_OD}
	}
\end{figure}

In addition, we analyzed the accuracy of detecting transmission values. The transmission information can be obtained from the modulation of the interference pattern. In the Methods section, it is explained how the modulation can be obtained from the sampled images. While doing so, we recorded phase shifting holograms with undetected light of an object in form of an happy face with different (but homogeneous) transmission values introduced by optical density (OD) filters.
An example for the obtained modulation image is shown in Fig.~\ref{fig:X_mod_vs_OD}(a) and (b) for OD 0 and OD 0.4, respectively. The modulation of the interference and the transmission of the object follow a proportional relationship~\cite{Gilaberte:2021}. The colored rectangles in Fig.~\ref{fig:X_mod_vs_OD}(a) mark three regions of interest used to verify the homogeneity of the calculated object transmission across different areas of the FOV. %used to quantitatively study the impact of different transmission conditions.
The dependency of the transmission on the inserted optical density filters for the three areas can be found in Fig.~\ref{fig:X_mod_vs_OD}(c) and agrees well with the theoretical prediction. 
In the Supplemental Material in Fig. \ref{fig:X_mod_noise} one finds the uncertainty, that is to be expected for a single measurement. For an exposure time of 500 ms and $M=4$ images used, one gets a noise level of $0.053\pm0.005$, which means the setup can detect a minimum transmission difference of 5\% to 6\%. The visible border around the eyes and mouth of the object are due to the limited spatial resolution (see Supplemental Material for a detailed explanation).

\section{Discussion}
%We have introduced a novel holography technique which retrieves the information of photons that are never detected.
We have introduced a novel technique to generate holograms with photons that are never detected. Our method is a quantum version of phase shifting holography, which retrieves full information, phase and amplitude, of an object. In order to do so, we make use of correlated signal and idler photon pairs, in a nonlinear interferometer configuration to obtain a hologram of the idler photon, through the measurement of the signal photon. Our technique is based on the quantum effect of 'induced coherence without induced emission', where information imprinted on the undetected (idler) photon can be retrieved analyzing the interference pattern of its partner (signal) photon.
%Thus the photon's information, transmission and phase, is retrieved to construct a hologram without a direct detection over the photon.\\
The main advantage of our technique over previous approaches is that we retrieved simultaneously the phase and amplitude of the shape of a photon in spite of never detecting this photon. 

Our approach also alleviates a key challenge towards integrating SU(1,1) interferometers in real-world biomedical applications: Previous demonstrations of imaging with undetected photons required long-term phase stability of relevant paths in an interferometer, and were thus either limited to operation in a controlled laboratory environment~\cite{Lemos:2014}, or involved the added technical complexity of active stabilization. Our approach has reduced the required level phase stability to the order of 2~s (the timescale over which a complete round of phase settings can be applied). Such levels of stability are straightforward to achieve in ruggedized optical assemblies~\cite{Gilaberte:2021} - a key advancement towards practical deployment. 

We quantify our method in terms of the achievable resolution given by the number of measurements and the acquisition time. It turns out that a combination of 500~ms acquisition time and four phase shifted images are an ideal combination in terms of recording time and image quality. We retrieve transmission values with a precision of down to 6\% and phases with a precision of down to $0.1\pi$.
Increasing the degree of spatial correlation between signal and idler as well as the numerical aperture of the imaging system can drastically enhance the spatial resolution of this method. Moreover, by combining our technique with quantum optical coherence tomography in nonlinear interferometers~\cite{Valles:2018,paterovaOCT2018,ramelowOCT2020,gerardOCT2020}, three-dimensional image reconstruction with undetected light becomes feasible.

%Is there a common metric for the sensitivity per intensity? Would it be possible to reduce further the image acquisition time? Can we say something on the phase resolution per intensity? How might this compare to a classical DPSH.

%\ST{ I may include some quantitative conclusion: DPSH is possible ... correct phase value with +- accuracy ... correct transmission value wit +- accuracy ... improvements can be done for spatial resolution}{}
%\mb{We demonstrated that obtaining the 2D hologram of an object is possible even through the interference of a single photon, instead of the interference between two photons recently demonstrated []}{}

\section{Materials and methods}
\subsection{Image Processing}
\begin{figure}[H]
	\centering
		\includegraphics[width=0.9\linewidth]{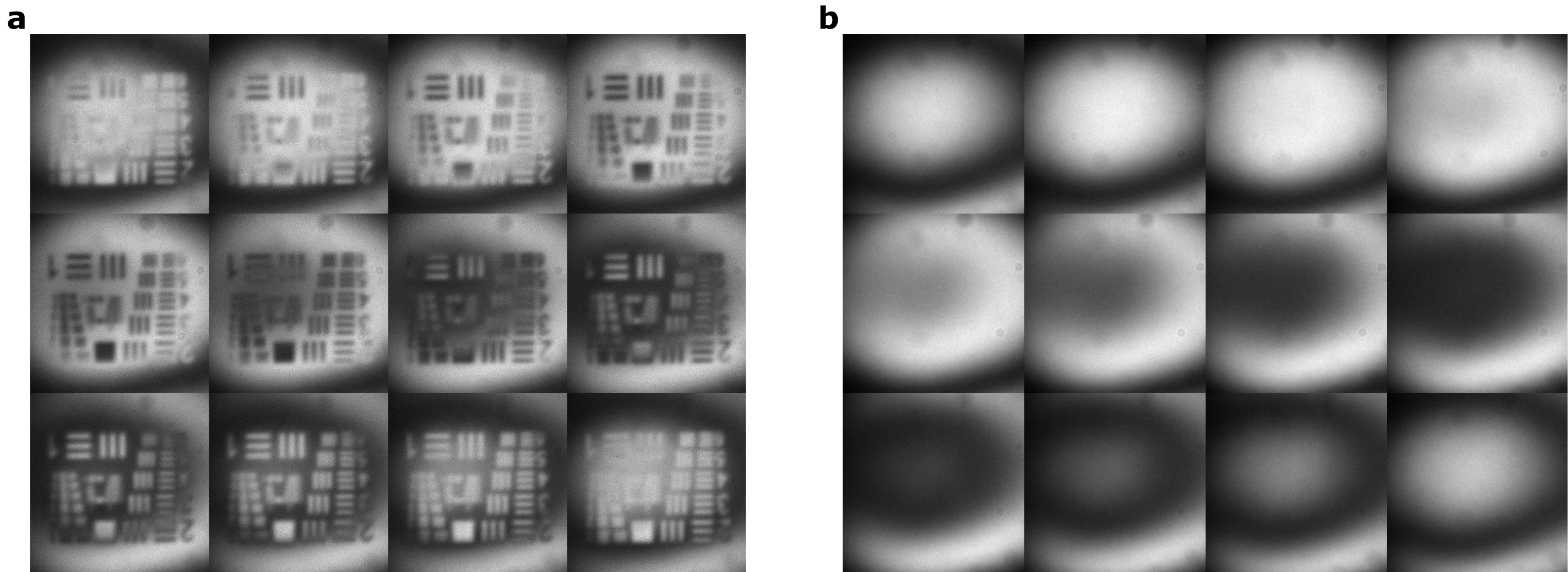}
	\caption{\textbf{Exemplary raw image set.} One set of recorded images for the phase shifting holography calculation ordered from left to right and top to bottom. (a) The object images. (b) The reference images.
	\label{fig:X_raw_imaages}
	}
\end{figure}

For the phase shifting holography calculation a set of twelve images of the pure interference as reference and another set of twelve images including the object is recorded (see Fig.~\ref{fig:X_raw_imaages}). To prepare the images for the calculation, they were filtered to reduce the noise level. The filtering has three sub-steps. First, the camera background is subtracted. As Background, a recording without pump beam was taken. Second, a 2D low-pass frequency filter is applied performing a Fourier transformation on the input images and cuts off high frequency noise. Last, a Gaussian filter is applied~\cite{scipy:2020}. With this the signal-to-noise ratio was increased by a factor of up to 1.82 depending on the exposure time and number of images.
\par
After the filtering, the phase and modulation are calculated. For this step, the images are separated into a subset depending on the algorithm to test. The base image set consist of twelve images taken at equally spaced phase positions with a phase difference of $\Delta\varphi_m=
%\frac{2m\pi}{M}
2m\pi/M ;m=0, ..., M-1$ to the first position ($m=0$). This set was split into four subset containing $M=12,6,4,3$ images, equally spaced inside the twelve from the base set and always starting at the same image $m=0$. This simulates measurements with less phase positions, without the need for additionally recordings to reduce differences between the measurements. 
\par
The calculation was done using a least-squares based algorithm for phase shift holography~\cite{malacara:2007}. The phase is obtained by Eq.~\eqref{equ:general_phase}, with $N_{\Delta\varphi_m}$ as the recorded intensity of image $m$.
%\begin{equation}
%    \theta = \arctan\left(\frac{-\sum N_{\Delta\varphi_m} \sin(\Delta\varphi_m)}{\sum N_{\Delta\varphi_m} \cos(\Delta\varphi_m)}\right)
%    \label{equ:X_phase_calculation}
%\end{equation}
A phase unwrapping algorithm is applied afterwards to resolve the periodicity~\cite{skimage:2020}. After this, the phase of the reference is subtracted to get the absolute phase change induced by placing the object inside the setup.
Our software gives real-time images (live feedback) for the expected modulation and phase. We introduced 300~ms delay time between each acquisition frame for the image post-processing. The live feedback uses a constantly updated circular buffer for the phase positions. The code for the final  post-processed image needs approx. 45~s to evaluate 360 raw images, i.e., 15 measurement sets containing 12 raw images of the reference and the object. The results are 15 phase and modulation images. The calculation time can be significant reduced implementing parallel-processing and graphics processing unit (GPU). This is also true for the live feedback system.
\par

In contrast to Eq.~\eqref{equ:general_amplitude}, losses at optical components will occur in the experiment. Hence, the interference modulation is measured, which is directly proportional to the transmission. Therefore, a referencing to an additional measurement without object is necessary to retrieve the absolute transmission values of an object. The modulation is calculated by Eq.~\eqref{equ:general_amplitude}. \\
%\begin{equation}
%    \gamma = \frac{N''}{N'} = \frac{2 \sqrt{\left( \sum N_{\Delta\varphi_m} \cos(\Delta\varphi_m) \right)^2 + \left( \sum N_{\Delta\varphi_m} \sin(\Delta\varphi_m) \right)^2}}{\sum N_{\Delta\varphi_m}}
%\end{equation}

%$N''$ is the interference amplitude and $N'$ is the average value of the intensity.
\subsection{Object fabrication}
The objects used in the experiments were generated by direct writing grayscale lithography. In doing so, a substrate is coated with a photoresist with a refractive index of 1.63 (at $\sim520$~nm) and exposed to UV-light in a particular pattern. The chemical solubility of the exposed areas increases in alkaline media and the micro-structure in the photoresist is formed.

\section{References}
\bibliographystyle{matthias_num}
\bibliography{Bib}

\section{Acknowledgements}

The authors want to thank Anja Schöneberg and Robert Leitel from Fraunhofer IOF for preparing the samples used in the measurements. \\

\subsection{Funding}
This work was supported as a Fraunhofer LIGHTHOUSE PROJECT (QUILT) and by the Fraunhofer Attract program (QCtech). Furthermore, funding support is acknowledged from the German Federal Ministry of Education and Research (BMBF) within the funding program Photonics Research Germany with contract number 13N15088. JPT acknowledges financial support from the Spanish Ministry of Economy and Competitiveness through the “Severo Ochoa” program for Centres of Excellence in R\&D (CEX2019-000910-S), from Fundació Privada Cellex, Fundació Mir-Puig, from Generalitat de Catalunya through the CERCA program, from project 20FUN02 “POLight” funded by the EMPIR programme co-financed  by  the  Participating  States  and  the  European  Union's  Horizon  2020  research  and innovation programme, and from project QUISPAMOL (PID2020-112670GB-I00) funded by MCIN/AEI /10.13039/501100011033. This project has received funding from the European Union's Horizon 2020 research and innovation programme under grant agreement No 899580.

%--------------------------------------
%\section{Authors information}
%--------------------------------------

%\subsection{Affiliations}
%\textbf{Fraunhofer Institute for Applied Optics and Precision Engineering IOF}

%\textit{Albert-Einstein-Str. 7, 07745, Jena, Germany} 

%Sebastian Töpfer, Marta Gilaberte Basset, Jorge Fuenzalida, Fabian Steinlechner, and Markus  Gräfe \\

%\textbf{ICFO-Institut de Ciencies Fotoniques, The Barcelona Institute of Science and Technology}

%\textit{08860 Castelldefels, Spain, and Dept. Signal Theory and Communications, Universitat Politecnica de Catalunya, 08034 Barcelona, Spain}

%Juan P. Torres \\

%\textbf{Friedrich-Schiller-University Jena, Abbe Center of Photonics}

%\textit{Albert-Einstein-Str. 6, 07745, Jena, Germany}

%Marta Gilaberte Basset, Fabian Steinlechner, and Markus  Gräfe \\
%-----------------------------------------------------
\subsection{Author Contributions}
M.G.B., F.S., and M.G. conceived the idea. J.P.T. provided the theoretical analysis. S.T., M.G.B., and J.F. designed and performed the experiments. S.T. conducted the computations and analyzed the data. All authors discussed the results and contributed equally in writing the paper and have given approval to the final version of the manuscript. \\
\subsection{Competing interests}
The authors declare that they have no competing interests. \\
\subsection{Data and materials availability}
All data needed to evaluate the conclusions in the paper are present in the paper and/or the Supplementary Materials.
%Correspondence to Markus Gräfe: markus.graefe@iof.fraunhofer.de
%
\subsection{Additonal information}
The work of this study is related to a pending patent application with the number EP20159989. The inventors are co-authors of this study.
%--------------------------------------------

% for the citation
\nocite{gatti2003}
\nocite{torres2011}
\nocite{boyd2008}

%No materials are reproduced from another source.
\end{document}